\documentclass{article}

\usepackage[english]{babel}

\usepackage[letterpaper,top=2cm,bottom=2cm,left=3cm,right=3cm,marginparwidth=1.75cm]{geometry}

\usepackage{amsmath}
\usepackage{graphicx}
\usepackage[colorlinks=true, allcolors=blue]{hyperref}

\newcommand{\aff}[1]{$^{\mathrm{#1}}$}

\title{Multi-Plane Light Conversion: A Practical Tutorial}

\author{
Yuanhang Zhang\aff{1}, Nicolas K. Fontaine\aff{2}}

\vspace{3mm}

\date{%
    \small{
    \aff{1}CREOL, the College of Optics and Photonics, University of Central Florida, Orlando, FL 32816, USA\\\aff{2}Nokia Bell Labs, 600 Mountain Ave., New Providence, NJ 07974, USA\\
    \textcolor{blue}{yuanhangzhang@knights.ucf.edu,   
      nicolas.fontaine@nokia-bell-labs.com}} \\
    \today
}

\begin{document}
\maketitle

\begin{abstract}
Multi-plane light conversion (MPLC) has recently been developed as a versatile tool for manipulating spatial distributions of the optical field through repeated phase modulations. An MPLC Device consists of a series of phase masks separated by free-space propagation. It can convert one  orthogonal set of beams into another orthogonal set through unitary transformation, which is useful for a number of applications. In telecommunication, for example, mode-division multiplexing (MDM) is a promising technology that will enable continued scaling of capacity by employing spatial modes of a single fiber. MPLC has shown great potential in MDM devices with ultra-wide bandwidth, low insertion loss (IL), low mode-dependent loss (MDL), and low crosstalk. The fundamentals of design, simulation, fabrication, and characterization of practical MPLC mode (de)multiplexers will be discussed in this tutorial.
\end{abstract}

\section{Introduction}
\subsection{MPLC: a historical review}

MPLC first emerged as a novel device that can realize unitary manipulation on light’s spatial profile, by using a finite number of phase masks separated by optical Fourier transform or free space propagation \cite{morizur2010programmable}. The authors of the very first paper of MPLC soon founded the company Cailabs that focuses on developing the beam-shaping technology. MPLC prototypes that support 3, 6, 10 and 15 modes were soon developed using 7, 7, 14, 20 phase masks, respectively \cite{labroille2014efficient,genevaux20156,labroille2016mode,barre2017broadband,bade2018fabrication}. The phase masks of MPLC can contain millions of pixels in total, which represent a huge parameter space for designers to control the wavefront of an optical beam. Considering the complexity of the MPLC system, Boucher \textit{et al.} applied the filtered random matrix (FRM) theory to show that, when considering high-dimensional modes outside the subspace of modes that MPLC is designed to shape, it behaves like a random scattering medium with limited number of controlled channels \cite{boucher2021full}. 
% Cailabs didn't disclose how many phase masks used in first several MPLCs. 

\begin{figure}[!t]
    \centering
    \includegraphics[width=5in]{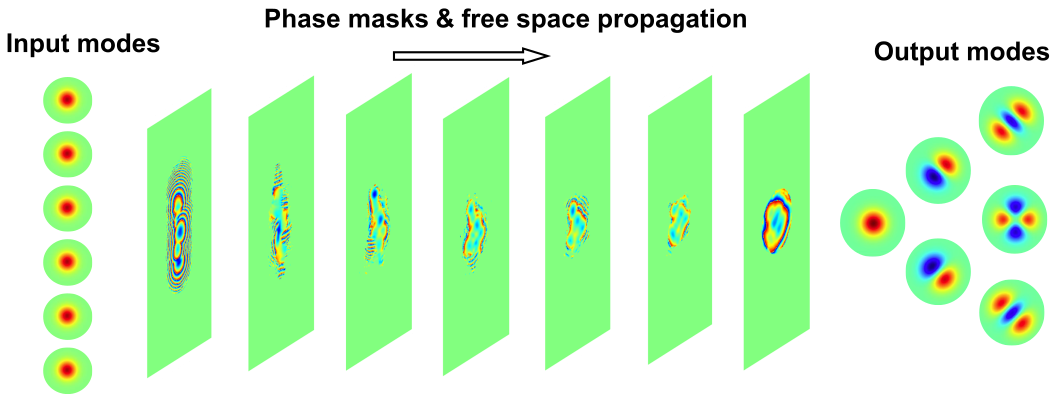}
    \caption{Basic structure of multi-plane light conversion. Input modes are orthogonal because they are spatially separated. Output modes are spatially overlapped but still orthogonal due to the inherent symmetries of these modes.}
    \label{mplc_intro1}
\end{figure}

For a real practical device, usually the input and output only contain a finite set of orthogonal spatial modes, which merely spans a subspace of the total mode space supported by the MPLC.  Within the finite set, modes are mutually incoherent. In telecommunication for example, as a mode (de)multiplexer, the inputs are spatially separated Gaussian spots while the outputs are spatially overlapped orthogonal mode sets, e.g. Hermite Gaussian (HG) or Laguerre Gaussian (LG) modes. These modes are more interested because of many good properties they possessed, e.g. HG modes are eigenmodes of the graded-index multimode fiber (GI-MMF) \cite{fontaine2018packaged}, and LG modes is more resistant to atmospheric turbulence \cite{belmonte2021optimal}. The basic structure of an MPLC that functions as a mode multiplexer is shown in Fig.\ref{mplc_intro1}.

Different methods have been used to optimize the patterns of phase masks. Since design of MPLC falls in the general category of inverse design, quite often, patterns of the optimized phase masks are not intuitive. Modified wavefront matching is the method commonly used to iteratively update phase masks \cite{sakamaki2007new}. Very like the famous Gerchberg–Saxton algorithm, it is not directly guided by any cost function but still converges fast through the gradient descent search. Another common method is to use algorithm differentiation on a cost function \cite{jurling2014applications}, for example, the gradient descent based backpropagation of neural networks \cite{lin2018all,zhu2022multiplane,barre2022inverse}. Regularization terms can be added to the cost function to smooth the mask, such as the well known total-variation (TV) norm \cite{beck2009fast}. Note that the multimode converters in \cite{barre2022inverse} uses hundreds of phase planes, which is more like a miniaturized version of the continuous, adiabatic photonic lantern rather than a free space device like MPLC. 

Generally transforming $N$ modes require an order of $N$ phase masks \cite{fontaine2022photonic}. But there is still no theory guides the minimum number of planes required for a specified transition. Actually, many factors affect the number of phase masks required to achieve a certain performance, such as the number of modes, the geometric layouts of the spots array at input, the mapping order, the chosen output mode sets, the pixel size, the mask plane spacing, etc.  In 2017, Fontaine \textit{et al.}  at Bell Labs found a magic mapping relation: From Cartesian points $(x,y)$ to the Cartesian indices $(m, n)$ of HG modes, mode conversion can be realized using remarkably few planes of equally spaced phase masks \cite{fontaine2017design}. 
Since then, 210/325 modes using 7 phase masks \cite{fontaine2019laguerre} and 1035 modes using 14 phase masks \cite{fontaine2021hermite} have been demonstrated. The reason for reduction of phase masks was explained by the separable mode basis theory in \cite{bade2018fabrication}. 

Although MPLC has many applications beyond the area of optical communication, this tutorial is mainly focus on devices for the telecommunication. The terminology common to all communication devices appear in this tutorial have the same meaning as section II of ref\cite{fontaine2022photonic}. Within telecommunication, especially within mode-division multiplexing which will be briefly introduced in the next section, mode (de)multiplexers are the most critical device.  

The performance of a mode (de)multiplexer is usually characterized by the mode coupling matrix $T$. The crosstalk matrix $X$ is given by $T^{\dag}T$, where $T^{\dag}$ is the conjugate transpose of $T$. By cascading a pair of multiplexer and demultiplexer,  matrix $X$ is usually obtained by measuring power at all output ports when launching light into each input port. After recovering the amplitude and phase of outputs using methods like the off-axis digital holography \cite{fontaine2019laguerre}, the matrix $T$ is retrieved by projecting the actual output modes to a set of desired mode basis. Then we apply the singular value decomposition (SVD) \cite{miller2019waves}, 
\begin{equation}
T = U\Sigma V^*
\end{equation}
where $U$ and $V$ are unitary matrices, representing linear supposition of the input/output mode basis with lossless mode coupling. $\Sigma$ is a diagonal matrix with non-negative real numbers called singular values, representing loss propriety of the device without crosstalk. As a result, the SVD operation projects the modes of the (de)multiplexer to another mode basis that separates crosstalk and loss. A good understanding between the two can be found in the supplementary of \cite{fontaine2019laguerre}. With the singular values, we can define the insertion loss (IL) and mode-dependent loss (MDL) as below

\begin{equation}
IL = \frac{1}{N} \sum_{n=1}^{N}|s_n|^2
\end{equation}

\begin{equation}
MDL = \frac{max(|s_n|^2)}{min(|s_n|^2)}
\end{equation}
where $s_n$ are the singular values of the complex mode coupling matrix $T$. 

In terms of other MPLC-based devices for optical communication, in 2018, an MPLC-based mode scrambler that completely inverts the first 10 modes of a GI-MMF was demonstrated to reduce the computational load of DSP \cite{li2018design}. In 2019, a mode demultiplexing hybrid (MDH) was proposed to simplify the optical front end of mode-division multiplexing (MDM) receivers by combining the functionality of mode demultiplexing and optical 90$^\circ$ mixing \cite{wen2019mode}. In 2021, an MPLC based stokes vector direct detection receiver was reported to measure the spatial complex amplitude and coherence without using a reference beam \cite{dahl2021high}. The authors also contributed to the design of a mode coupler \cite{zhang2019slab}, a reconfigurable mode router \cite{zhang2020reconfigurable}, and an ultrabroadband optical hybrid \cite{zhang2020ultra} using the concept of MPLC. MPLC has also been reported to combat the atmosphere turbulence, either using the static scheme \cite{fontaine2019digital,billaud2022turbulence} or the dynamic scheme \cite{alsaigh2023bulk}. 

Based on the space-time duality: diffraction in space is similar to dispersion in time, the concept of MPLC has also been extended to temporal domain for waveform shaping. Mazur \textit{et al.} used a series of phase modulators separated by dispersive all-pass filters to fulfill a lossless optical arbitrary waveform generator \cite{mazur2019multi}. Ashby \textit{et al.} proposed a similar scheme for applications in quantum information science, which will be favored by the quantum community because of the lossless operation inherited in unitary transformation on temporal modes \cite{ashby2020temporal}. Combined with a wavelength selective switch (WSS), MPLC has also been used to control both the spatial and temporal properties of light at the same time \cite{mounaix2020time,mazur2020gain}. MPLC is also a versatile tool to study the spatiotemporal properties of light \cite{cruz2022synthesis,shen2022roadmap}.

In 2018, Lin \textit{et al.} trained the neural networks for all-optical machine learning, which are formed by several transmissive phase masks in a similar structure to MPLC \cite{lin2018all}. Since then, the potential of using MPLC to do matrix multiplication at light speed has raised much attention \cite{zhou2022photonic,li2019class,bernstein2021freely}, such as the research on photonic Ising machine \cite{pierangeli2019large} and the neuromorphic optoelectronic computing \cite{zhou2021large}. In 2022, Ozer \textit{et al.} reported using an MPLC-based mode sorter for super-resolution imaging \cite{ozer2022reconfigurable}.

For small volume or integrated MPLC, besides the millimeter-sized one mentioned above \cite{barre2022inverse}, a 3-mode MPLC using the metasurface design was reported in 2020 \cite{oh2020compact}. Mazur \textit{et al.} demonstrated 36- and 45-mode MPLCs with a reduced size \cite{mazur2019reduced}. Pastor \textit{et al.} proposed to used arrays of phase shifters and multimode interference (MMI) waveguides to implement arbitrary unitary transformation\cite{pastor2021arbitrary}.  Researchers from Yoshiaki Nakano's group in Japan are developing integrated reconfigurable optical unitary converters (OUCs), and found OUCs based on the MPLC structure is more robust to fabrication errors than the traditional Mach-Zehnder interferometer (MZI)-based OUC \cite{tang2017integrated,tang2018reconfigurable,tanomura2022scalable}. 

Against from the trend of trying to minimize the structure, the other attempt is designing large-area MPLCs which could be used for high power applications. Cailabs has developed the CANUNDA solution for laser machining by shaping high-power continuous or pulsed beam \cite{meunier2020fully}. At CREOL there is on-going research of using large-area non-mode-selective MPLC \cite{wen2021scalable} for high-power wavefront synthesis, which could be used in applications such as long distance free space communication in the turbulent atmosphere.

\subsection{Mode-division multiplexing in optical communication}

\begin{figure}[!h]
    \centering
    \includegraphics[width=3in]{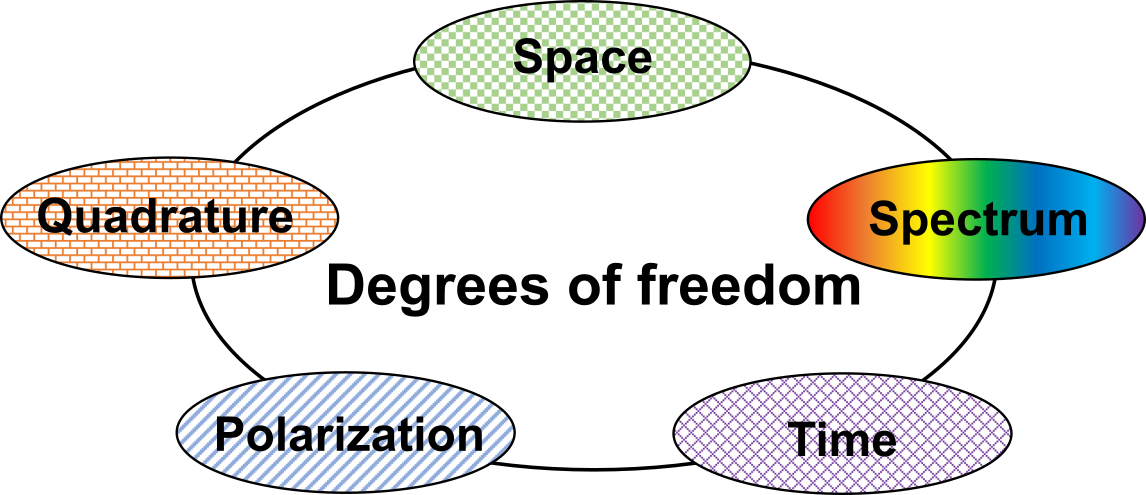}
    \caption{Degrees of freedom for capacity scaling in optical fiber communication systems.}
    \label{mplc_intro2}
\end{figure}

Optical fiber communication, as the backbone infrastructure of internet, has been supporting our increasingly information driven society for many decades. So far much of this progress has been in exploring the data-carrying capacity of a single mode fiber (SMF). To achieve this goal, researchers have attempted to multiplex information into time, frequency, polarization, quadrature (amplitude and phase) of an optical field, as shown in Fig.\ref{mplc_intro2}. In all these methods, most data traffic demand has been met by the wavelength-division multiplexing (WDM) technology, which increases the spectral bandwidth of fiber-optic communication channel by 2 orders of magnitude in the past two decades. The deployment of broadband, low noise Erbium doped fiber amplifiers (EDFAs) enables amplification of hundreds of wavelength channels simultaneously. Polarization multiplexing has been used to double the transmission capacity. In addition, the reviving coherent optical communication system enables use of advanced modulation formats (e.g., quadrature phase-shift keying (QPSK), quadrature amplitude modulation (16 QAM, 64 QAM, etc.) in the entire amplitude-phase complex plane. With the coherent detection of the complex electrical field, phase and polarization management, dispersion and nonlinear compensation are available by the digital signal processing (DSP) \cite{li2009recent}. Actually, the spectral bandwidth of the fiber-optic communication channel can be further increased by exploiting the low-loss transmission window of optical fiber beyond the C bands ($\sim$5THz, 1530 – 1565 nm) by using the dry fibers \cite{essiambre2010capacity} or the hollow-core fibers, which promises roughly 10$\times$ capability of increase if using the entire O through L bands ($\sim$50 THz, 1260 – 1625 nm). However, to meet the exponentially growing data traffic, we need to dig out more capacity from the optical fiber. In the information theory, Shannon's equation gives the maximum capacity that can be reliably communicated over a noisy channel \cite{shannon1948mathematical,winzer2018fiber}:
\begin{equation}
    C = M\times B\times 2 \times log_2(1+SNR)
\end{equation}
where $SNR$ is the signal-to-noise ratio, the factor 2 is the degree of freedom of polarization, B is the system bandwidth, and M is the number of spatial channels. As the capacities provided by the mature technology of WDM and coherent communication quickly approaching their fundamental Shannon limits, the only untapped dimension in single fibers is the spatial dimension. As a "pre-log" factor in the above equation, it is expected to provide potentially unbounded scaling of capacity \cite{li2014space,richardson2013space}. Multi-core fiber (MCF) and multi-mode fiber (MMF) are two most promising fibers supporting space-division multiplexing (SDM). And MCF is usually thought as a transition technology from current SMF to the ultimate goal of MMF, because MMF provides the highest density of spatial channels \cite{ryf2016space}. As a result, mode-division multiplexing (MDM) comes to the forefront of the optical communication research.

% add the Shannon capacity equation. modes and bandwidth are pre-log factor!

Mode (de)multiplexers are the necessary components to interface the transceivers with the MMF. Of all the mode (de)multiplexing devices, MPLC stands out because it has shown great potential in designing devices with ultrabroadband, low insertion loss (IL), low mode-dependent loss (MDL), low crosstalk, in addition to scaling to large number of modes and large-scale fabrication capability. In 2021, Peta-bit transmission has been demonstrated using a 15-mode MPLC \cite{rademacher2021peta}. The record was soon overtaken by a 1.53 Peta-bit/s transmission using 55 modes by the same group in 2022 \cite{rademacher20221}.

\section{MPLC-based mode (de)multiplexers}
In this section, we explain the fundamentals of design, simulation, fabrication, and characterization of practical MPLC mode (de)multiplexers.

\subsection{Simulation}

\subsubsection{Broadband, reflective MPLC with tilted angles}

\begin{figure}[h]
    \centering
    \includegraphics[width=6in]{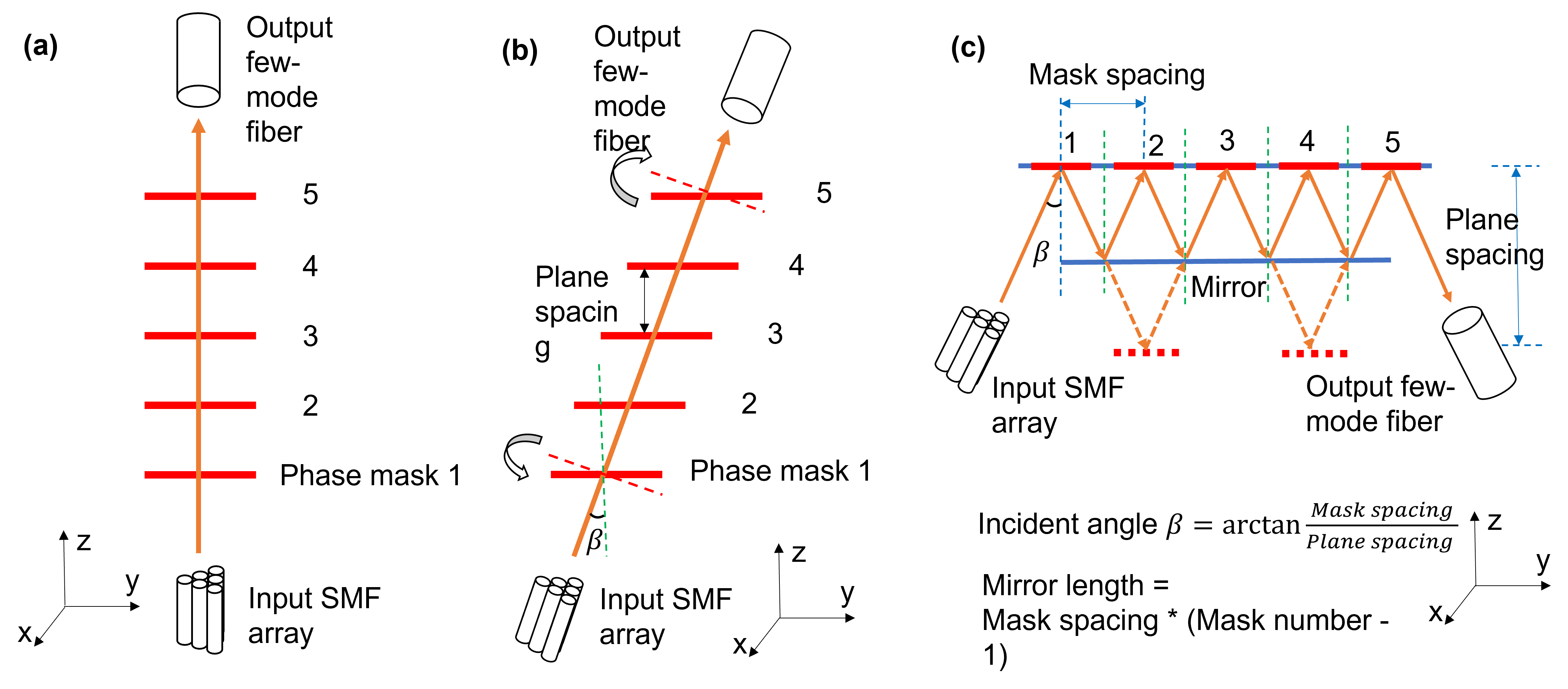}
    \caption{(a) In the transmissive MPLC, the incident beam is normal to the phase masks. Traditional angular spectrum method (ASM) is used in this architecture. (b) Transmissive MPLC where the beam has an incident angle to the phase masks. This architecture serves as an intermediate step to understand the reflective MPLC in (c). The translated ASM method is used for free space propagation. (c) The reflective MPLC with a folded cavity consists of a phase mask plane and a mirror. }
    \label{mplc_funda_1}
\end{figure}

A simple MPLC design starts with the transmissive architecture where the incident beam is normal to the phase masks as shown in Fig.\ref{mplc_funda_1}(a), whereas a practical MPLC usually comes in the form of a folded cavity including a phase mask plane and a mirror plane, as shown in Fig.\ref{mplc_funda_1}(c). The folded cavity simplifies the alignment and assembling of a real MPLC device. In Fig.\ref{mplc_funda_1}(a), traditional angular spectrum method (ASM) is used for free space propagation. However, traditional ASM has a constraint in simulating diffraction: The incident beam must be normal to the input and output plane. i.e., no tilted planes are allowed. It also requires the input sampling window and the output sampling window have the same size and center, which increases the computational load dramatically when the output plane is translated horizontally, as this sampling window has to be increased to be able to contain both the input field as well as the output field. Therefore, traditional ASM limits the ability to simulate a real MPLC device with an incident angle.

\begin{figure}[b]
    \centering
    \includegraphics[width=5.5 in]{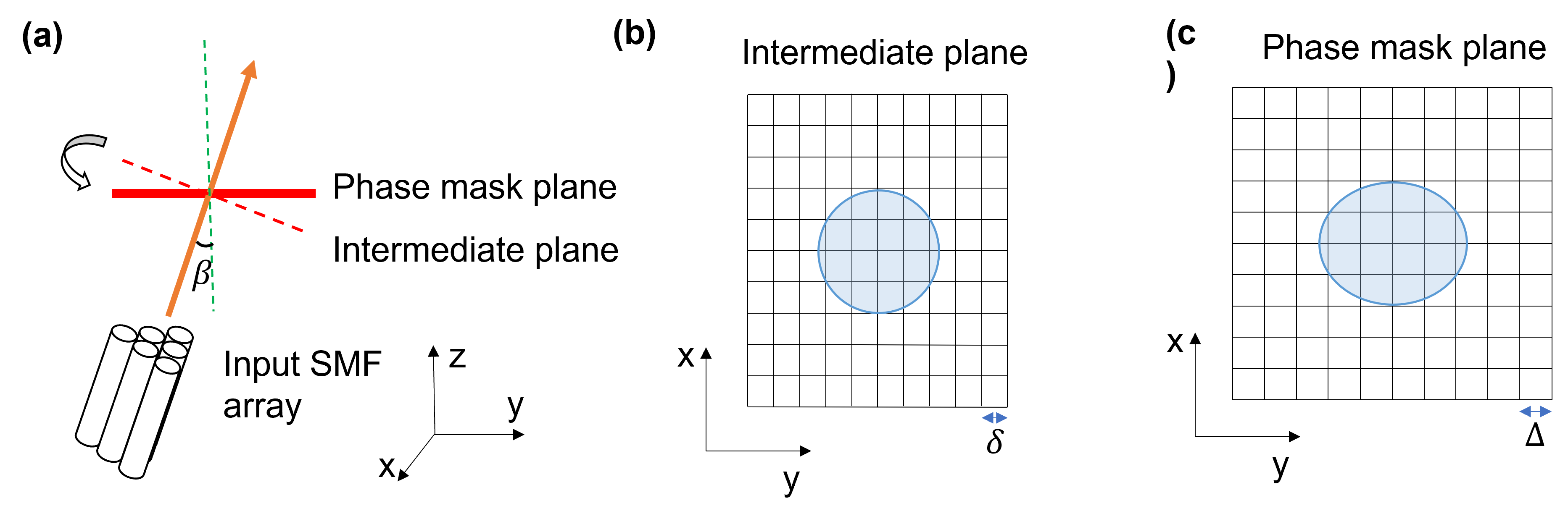}
    \caption{(a) The beam coming out of the collimated input fiber array has a wavefront normal to the optical axis. But the phase masks are parallel to the y-axis. For small incident angle (usually less than 10 degrees) we can simply re-scale the sampling grid in the $y$ direction. (b) and (c) show the sampling grid in the intermediate plane and phase mask plane, respectively.  Their grid size is same in the $x$ direction but is related by $\delta = \Delta  cos(\beta)$ in the $y$ direction.}
    \label{mplc_funda_2}
\end{figure}

Before explaining the reflective design of Fig.\ref{mplc_funda_1}(c), we use Fig.\ref{mplc_funda_1}(b) as an intermediate step to help understand the process.  In Fig.\ref{mplc_funda_1}(b) we build a coordinate system where the incident beam has a tilted angle relative to the $z$ direction. The translated ASM \cite{son2015light,ritter2014modified} is applied to simulate beam propagation between horizontally shifted planes. The translated ASM has a similar numerical complexity as the traditional ASM. It takes the $k$ vector phase term in $y$ direction out of the Fourier integral as a carrier frequency and let the sampling window shifts together with the phase mask. This helps relieve the sampling requirement in the calculation. However, it should be noted that the collimated beam coming out of the input single-mode fiber array or the beam going into the output few-mode fiber has a wavefront normal to the optical axis. So, a rotation is necessary before multiplying the phase pattern of the first phase mask at the input or after the last phase mask at the output, respectively. One calculation method for optical diffraction on tilted planes is to use the coordinate rotation in Fourier domain \cite{matsushima2003fast}. However, this is based on a nonuniform sampling interval in the Fourier plane and requires two-dimensional (2D) interpolation, which has created serious computational complexities. Readers interested in details of this method can refer to chapter 9 of Dr. Matsuhima’ s new book \cite{matsushima2020introduction}. For small incident angle (less than 10 degree, which is the common case for a MPLC device) however, we can simply rescale the $y$ coordinate by $cos(\beta)$, which will be talked in detail in Fig.\ref{mplc_funda_2}. In certain special cases, if we know the analytic equation of the input/output beam (e.g., Gaussian beam, HG modes, LG modes), the field distribution cut by a titled field can be written directly with coordinates rotation. In the configuration of Fig.\ref{mplc_funda_1}(c), the incident angle is determined by the mask spacing and the plane spacing. The mask-spacing is the distance between two adjacent masks on the mask plane, which is usually same as or slightly larger than the mask width in the real design. The mirror width should be diced to the exact length for not blocking the input/output beam, which is the product of the mask-spacing and (mask number-1). The height of the mirror can be oversized to collect more diffracted beam. Note that instead of Fig.\ref{mplc_funda_1}(c), Fig.\ref{mplc_funda_1}(b) is how the computer program understands the physical structure. The shared MATLAB code in \cite{fontaine2019laguerre} gives the basic example code of the structure in Fig. \ref{mplc_funda_1}(a), which is a good starting point to design reflective MPLCs.

Figure \ref{mplc_funda_2} shows the approximation that the field on a tilted plane can be calculated by re-scaling the sampling grid in the $y$ direction. The re-scaling factor is given by $cos(\beta)$. After re-scaling, a perfect circle (resembling the Gaussian mode) in the intermediate plane is stretched to an ellipse in the phase mask plane, as shown in Fig.\ref{mplc_funda_2}(b) and (c). We choose to use equidistant sampling grids (square grids) for the phase mask calculation because the fabricated phase masks or phase-only liquid crystal on silicon (LCoS) devices have square pixels. As a result, the sampling grids on the intermediate plane have a rectangular shape.

It's ideal to consider a broadband design at the beginning to generate very smooth phase masks. This also brings the benefit of improving the tolerance to misalignment. Introducing broadband design can be realized by simply modifying the original code of updating phase masks. The trick is besides taking the average phase differences of the forward and backward field of all spatial modes, we can add another \textbf{$for$} loop to average at several wavelengths close to the center wavelength. Besides the multi-wavelength optimization, ref\cite{fontaine2017design} also provides other suggestions to smooth phase masks. The “equivalent” function of wavelength and mode leads to another design from the author -- a wavelength-mode sorter \cite{zhang2020simultaneous}, which may find applications in spectral/wavelength beam combining. That design was later experimentally verified by a group at Shenzhen University \cite{kong2021mode}.

\subsubsection{Sampling requirements}

The transfer function of free space, or the kernel of the angular spectrum method (ASM), is given by \cite{saleh2019fundamentals}
\begin{equation}
    H(v_x,v_y) = exp(-j2\pi d\sqrt{\lambda ^{-2}-v_x^2-v_y^2}) = exp[-j\phi(v_x,v_y)]
\end{equation}
with $\phi(v_x,v_y) = 2\pi d\sqrt{\lambda^{-2}-v_x^2-v_y^2}$. After reorganizing the term, we got an ellipsoid function:

\begin{equation}
    \frac{\phi^2} {(\frac{2\pi d}{\lambda})^2} + \frac{v_x^2}{\lambda ^{-2}} + \frac{v_y^2}{\lambda ^{-2}} = 1
    \label{ellispoid}
\end{equation}

\begin{figure}[!b]
    \centering
    \includegraphics[width=4.6in]{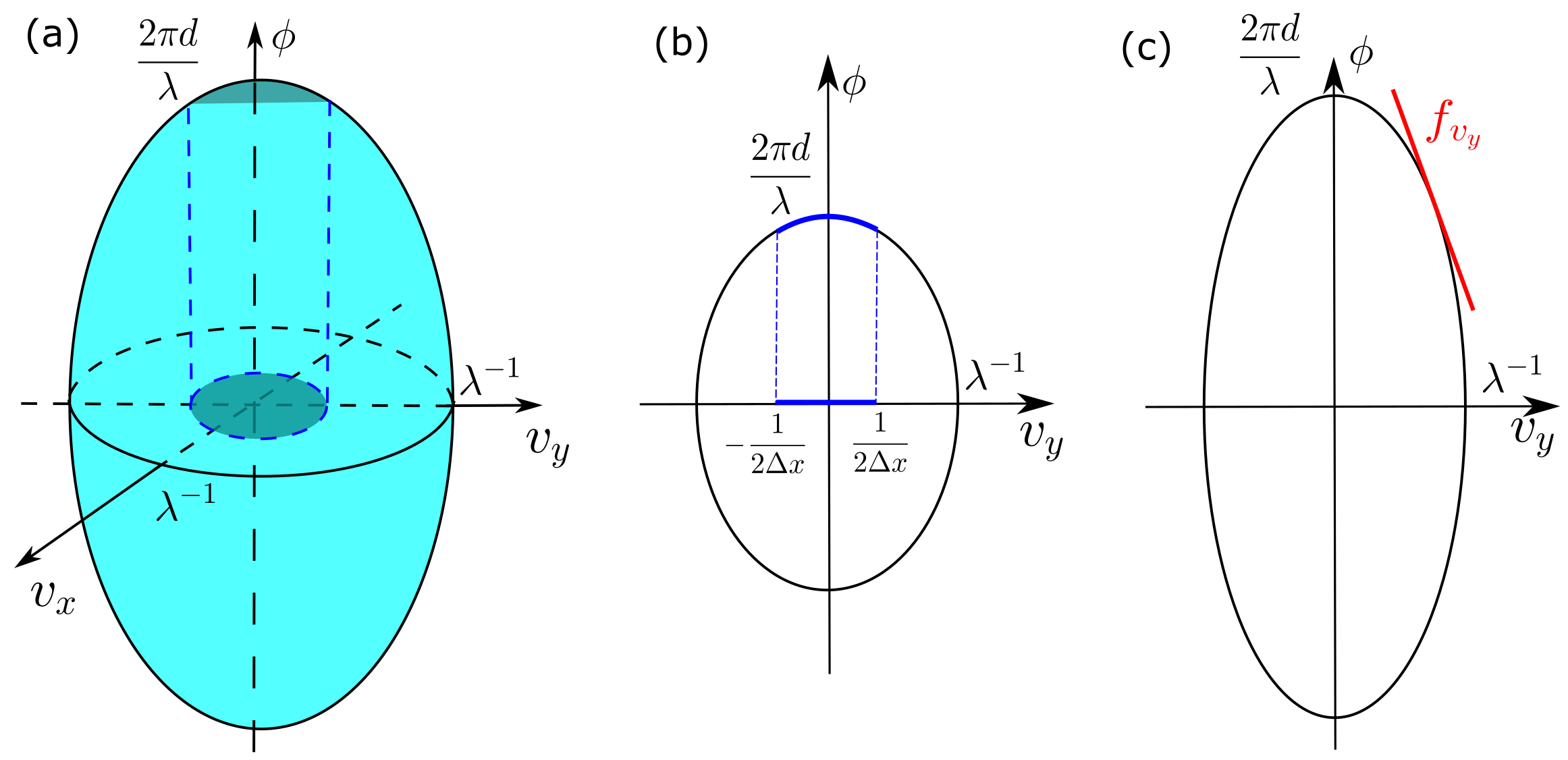}
    \caption{ (a) Phase of the transfer function of the ASM is an ellipsoid. Cut in $\phi-v_y$ surface with (b) a small $d$ and (c) a large $d$. Local spatial frequency is the slope of the ellipsoid indicated by the red line in (c). For paraxial beam ($v_x,v_y \ll \lambda^{-1}$), the transfer function can be approximated by a parabolic function (indicated by the gray region in (a) and blue curve in (b)). }
    \label{mplc_funda_3}
\end{figure}

The above ellipsoid equation is shown in Fig.\ref{mplc_funda_3}(a).

In the digitized version, the sampling points in the spatial domain is
\begin{equation}
\begin{split}
    x &= (-\frac{N_x}{2}:\frac{N_x}{2}-1) \Delta x   \\
    y &= (-\frac{N_y}{2}:\frac{N_y}{2}-1) \Delta y
\end{split}
\end{equation}

where $\Delta x$ and $\Delta y$ are the grid size, or pixel size for the fabricated phase masks or LCoS. Without loss of generality, we assume $\Delta x = \Delta y$ here. $N_x$ and $N_y$ are the pixel numbers in $x$ and $y$ direction, respectively. The corresponding sampling points in the Fourier domain (or k-space, with $k_x = 2\pi v_x, k_y = 2\pi v_y$) is
\begin{equation}
\begin{split}
    v_x &= (-\frac{N_x}{2}:\frac{N_x}{2}-1) \Delta v_x   \\
    v_y &= (-\frac{N_y}{2}:\frac{N_y}{2}-1) \Delta v_y
\end{split}
\end{equation}
where $\Delta v_x = 1/(N_x \Delta x)$ and $\Delta v_y = 1/(N_y \Delta y)$ are the grid size in the Fourier domain. The spatial frequency range of $v_x$ is $-\frac{1}{2\Delta x}$ to $ (\frac{1}{2\Delta x}-\frac{1}{N_x\Delta x}) \approx \frac{1}{2\Delta x}$. The sampling range in k-space is inversely proportional to the pixel size, but not related to the pixel number. To avoid aliasing, sampling points (pixel number) $N_x$ and $N_y$ must be large enough to resolve the finest structures in the optical field. 

Note that the foot of the ellipsoid is always fixed at $\lambda ^{-1}$, while the height is related to the propagation distance $d$. As $d$ increases, the slope increases dramatically which puts a more strict constraint for the sampling requirements. When $d$ is large enough, it will always introduce aliasing without limiting the bandwidth in the ASM method. In Dr. Matsushima's digital holography book \cite{matsushima2020introduction}, a band-limited ASM method is discussed by adjusting the cutoff frequency as propagation distance changes. 

As shown in Fig.\ref{mplc_funda_3}(c), the slope of the ellipsoid gives the "local spatial frequency", which is shown below and is proportional to the distance $d$. 
\begin{equation}
\begin{split}
    f_{v_x} &= \frac{1}{2\pi}\frac{\partial \phi(v_x,v_y)}{\partial v_x} = \frac{v_x d}{\sqrt{\lambda ^{-2}-v_x^2-v_y^2}}    \\
    f_{v_y} &= \frac{1}{2\pi}\frac{\partial \phi(v_x,v_y)}{\partial v_y} = \frac{v_y d}{\sqrt{\lambda ^{-2}-v_x^2-v_y^2}}
\end{split}
\end{equation}
Considering $f_{v_x}$ for example, the maximum of local spatial frequency is given by
\begin{equation}
    max(f_{v_x}) = \frac{max(v_x)d}{\sqrt{\lambda ^{-2}-max(v_x)^2-max(v_y)^2}}
\end{equation}
If we take $max(v_x)=(N_x/2-1)\Delta v_x \approx 1/(2\Delta x)$ and $max(v_y) \approx 1/(2\Delta y)$ into above equation, and considering the Nyquist sampling condition (sampling frequency should be at least twice of the highest frequency of the signal), i.e., 
\begin{equation}
\frac {1}{\Delta v_x} > 2 max(f_{v_x}) 
\end{equation}

we will get
\begin{equation} \label{condition1}
 d < \frac{\Delta x^2}{2}N_x\sqrt{\frac{4}{\lambda ^2} -\frac{1}{\Delta x^2} -\frac{1}{\Delta y^2}}
\end{equation}

For MPLC with a pixel size of $\Delta x = \Delta y = 8 \mu m, N_x =N_y=512, \lambda =1550 nm, d < 21 mm.$

For a paraxial beam, $\frac{1}{2\Delta x} \ll \lambda ^{-1}$. The ratio between these two gives us a sense if we are in the paraxial domain. In order to make sure the paraxial beam condition (also known as Fresnel approximation or parabolic approximation) is satisfied, we expand the phase term $\phi$ by Taylor expansion
\begin{equation}
    \phi(v_x,v_y) = 2\pi d\sqrt{\lambda ^{-2}-v_x^2-v_y^2} = \frac{2\pi d}{\lambda}[1-\frac{v_x^2+v_y^2}{2\lambda ^{-2}}+\frac{1}{8}(\frac{v_x^2+v_y^2}{\lambda ^{-2}})^2-\cdots]
\end{equation}

For the parabolic approximation to be valid, the third phase term should be much less than $\pi$, i.e.,
\begin{equation} \label{Fresnel_condition}
    \frac{2\pi d}{8\lambda}(\frac{v_x^2+v_y^2}{\lambda ^{-2}})^2 \ll \pi
\end{equation}
which gives another constraint on $d$ if we use $max(v_x) = 1/(2\Delta x), max(v_y) = 1/(2\Delta y)$ and $\Delta x = \Delta y$,
\begin{equation} \label{condition2}
    d \ll \frac{16\Delta x^4}{\lambda ^3}
\end{equation}

For the MPLC with $\Delta x = \Delta y = 8 \mu m$ at 1550 nm, $d \ll 17.6$ mm. 

Note that conditions Eq.(\ref{condition1}) and Eq.(\ref{condition2}) is sometimes too strict as it requires the entire sampling range to satisfy the Nyquist sampling condition or parabolic approximation while in reality the signal bandwidth is much smaller than the sampling frequency range. The pixel number $N_x$ and $N_y$ can also be very large at the beginning (pad with zeros for the computational window), and the phase mask can be cropped to a smaller size after optimization. In summary, the pixel number and pixel size should be chosen carefully in order to avoid aliasing issues. The pixel size determines the sampled range of the spatial frequency of the angular spectrum method (ASM), and the pixel number determines whether we have enough sampling points to get rid of aliasing. 

\subsubsection{Considerations for fabrication-ready design}
In this section, design of a 45-mode mode selective MPLC with a linear input array was used as an example. Although the specific configuration that inputs arranged in a right triangle shape and outputs using Hermite-Gaussian(HG) modes has been reported as an effective way to reduce the required number of phase masks, using a standard one-dimensional (1D) linear fiber array instead, could potentially reduce the cost of final product and make the alignment much easier.

\begin{figure}[h]
    \centering
    \includegraphics[width=6in]{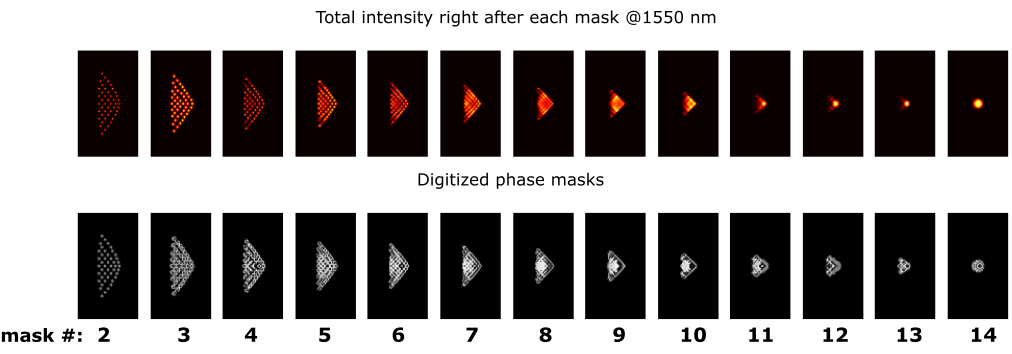}
    \caption{Initialization phase of the 45-mode MPLC design, using 13 phase masks (2nd to 14th).}
    \label{mplc_funda_4}
\end{figure}

This MPLC use 14 phase masks to convert a linear input Gaussian spots to HG modes. Input and output beam waist diameter are 70 $\mu m$ and 280 $\mu m$, respectively. The beam waist to the first and last mask spacing are 5 mm and 20 mm, respectively. The mask-mirror spacing is 10 mm with an incident angle 7.29 degree. Each mask has a pixel number of 1200 by 512 and a pixel size of 5 $\mu m$.

Because the mapping order from the right triangle to the HG modes is already known to reduce the number of phase masks \cite{fontaine2019laguerre}, the mapping order from the linear array to the triangle array should also be optimized to effectively use the first 7 phase masks. Ideally, total deflection angles of all beams should be minimized to encourage smooth phase mask solutions. In my design, I found directly optimizing from a linear array to the HG modes using the wavefront matching method doesn't give good results. As a result, I optimized the second to the last mask first with input positions of each beam manually set for the first mask, as shown in Fig. \ref{mplc_funda_4}. For the first mask, the spots layout should span an area as large as possible (which was found helpful to reduce MDL) and has a shape between the linear array and the right triangle array. 

\begin{figure}[h]
    \centering
    \includegraphics[width=6in]{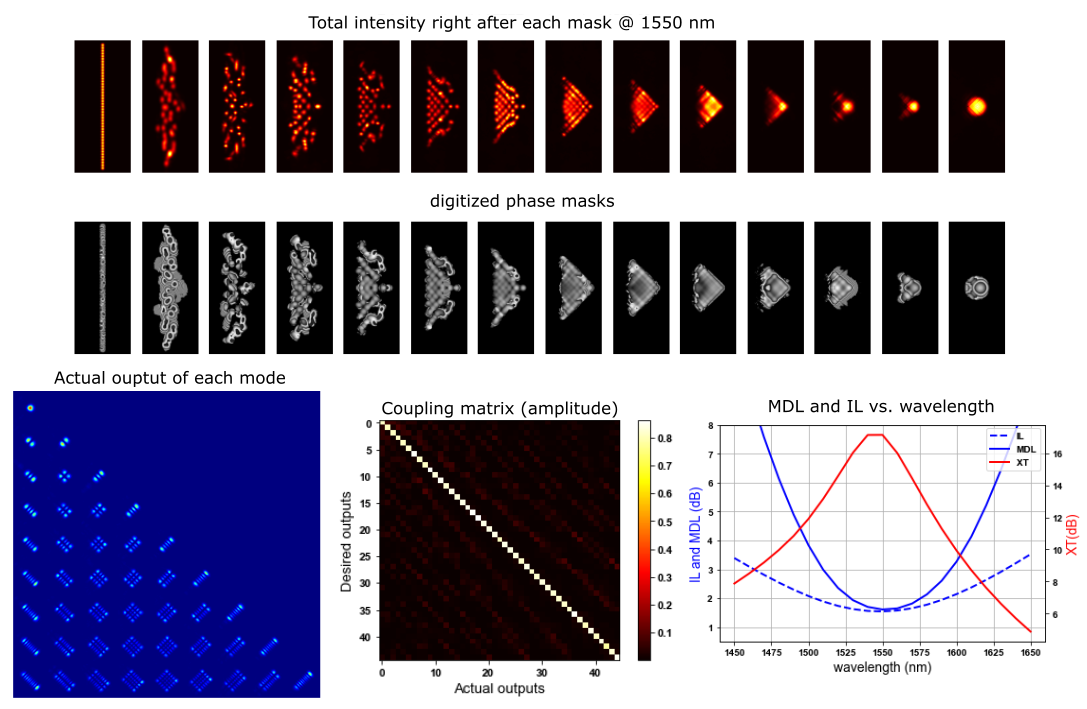}
    \caption{Final results of the 45-mode MPLC with a linear input array.}
    \label{mplc_funda_5}
\end{figure}

After this procedure, these masks were used as the initial state of the entire optimization. The results are shown below in Fig. \ref{mplc_funda_5}. Note the pixel number of each mask was also extended to 1800 by 1024 by padding zeros in the optimization to avoid aliasing. After convergence, it was cropped to 1200 by 512 to save space on the wafer.

To fabricate the reflective phase mask using the multi-step binary lithography method, we need to etch multiple steps to give digitized phase levels between 0 and $2\pi$ phase shift.  For example, digitizing to 64 levels needs 6 steps as $2^6=64$. 2$\pi$ corresponds to 775 nm etch depth at 1550 nm (Recently, MPLC with 8$\pi$ phase wraps was found to have better performance than their 2$\pi$ counterparts using less phase masks \cite{fontaine2022broadband}). After using the $angle()$ function extracting the phase term $\Phi$, by default the phase is wrapped between $-\pi$ and $\pi$ and the background is 0. Directly mapping the phase to the etching depth as shown in Fig.\ref{mplc_lithography}(a) is usually not desired for fabrication, as the factory (SILIOS Technologies, Holo/Or, etc.) don't want to etch the background. This can be simply solved by a trick,
\begin{equation}
\begin{split}
    \Phi  &= \Phi +\pi       \\
    \Phi[\Phi \geq \pi] &= \Phi[\Phi \geq \pi] - 2\pi
\end{split}
\end{equation}
After the above conversion, the background, originally with a phase 0 (corresponding to an etch depth of 775/2 nm), is shifted to $-\pi$ (corresponding to an etch depth of 0).

\subsection{Experiment}

\subsubsection{Make a collimated linear fiber array}
\begin{figure} [h]
    \centering
    \includegraphics[width=5.6in]{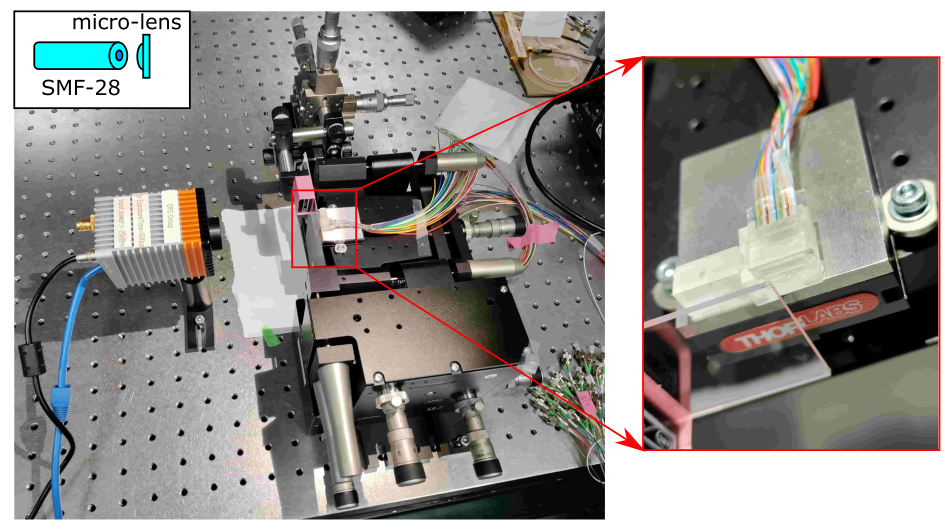}
    \caption{Assembling a collimated linear fiber array.}
    \label{mplc_funda_6}
\end{figure}

A six-axis stage with sub-micron resolution (Thorlabs, MAX601D/M - 6-Axis NanoMax Stage) was used to assemble the collimated linear fiber array. Usually the fiber array with a standard pitch (127 $\mu m$ or 250 $\mu m$) was used. The microlens array (SUSS MicroOptics) with the same pitch and anti-reflection coating between 1520 nm and 1620 nm was used to collimate the beam.  A plano-convex lens is usually used to collimate a beam with the flat surface facing the fiber tip. However, in the experiment shown in Fig.\ref{mplc_funda_6}, the microlens array we are using (pitch = 127 $\mu m$, size 6.658 mm $\times$ 3.229 mm, radius of curvature 0.155 mm) has a thick fused silica substrate. The focal length is so short that the focal point on the flat surface side is buried inside the substrate, so we have to use the convex surface facing the fiber array. We glue the corner of the flat surface of the microlens array (with a tiny amount of glue) to a glass plate first and then bring it to the V-grove fiber array. After proper alignment all ports should have a perfect Gaussian shape and the beam waist radius was measured to be close to 35 $\mu m$. Then we use UV glue to fix the microlens array to the fiber array and pull the glass plate in backward direction to detach it. A good collimated fiber array has similar collimated beam size for all ports with small pointing errors.

\subsubsection{Retro-reflection in alignment}

\begin{figure}[!h]
    \centering
    \includegraphics[width=6in]{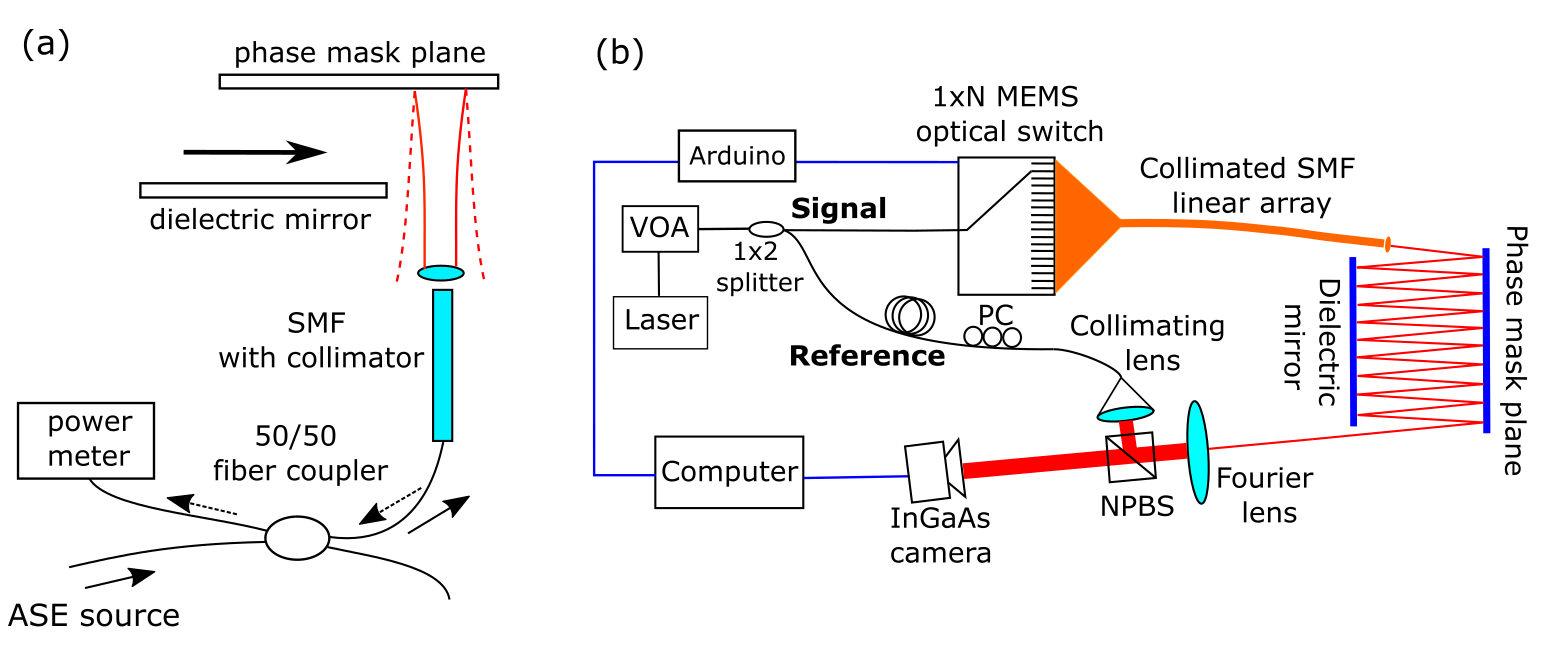}
    \caption{(a)In alignment, retro-reflection was used to align the mirror and fiber array.(b) Experimental setup to characterize a MPLC. VOA: variable optical attenuator, NPBS: non-polarization beam splitter, PC: polarization controller.}
    \label{mplc_funda_7}
\end{figure}

The phase mask plane is glued on a glass base first. The dielectric mirror and input fiber array are held on a 6-axis (x,y,z,pitch, yaw, roll) stage respectively. As shown in Fig.\ref{mplc_funda_7}(a), an ASE source, a 50:50 fiber coupler, a power meter and a collimated SMF are used to assist the alignment.  The alignment takes 3 times of retro-reflection. First, the dielectric mirror was moved away and we do retro-reflection on the empty region on the phase mask. We adjust the stage holding the SMF collimator until maximum power reading on the power meter. In this way, we make the beam normal to the phase mask plane. Next the dielectric is moved in and do retro-reflection on the mirror. This time we adjust the 6-axis stage holding the mirror until maximum power reading. After this step we know the mirror is parallel to the mask plane. Finally, we do retro-reflection with the input fiber array to make it normal to the mask plane first. Then we rotate the fiber array to the correct incident angle. 

Fig.\ref{mplc_funda_7}(b) shows the experimental setup to characterize the MPLC using off-axis digital holography. The 1$\times$N optical switch is used to selectively launch beam to all output HG modes. The optical switch is synchronized with the InGaAs camera so that whenever the input port changes, a frame of raw image is captured. A Fourier lens (2f system) was put at the output of MPLC, and it relays the flat wavefront of the beam waist to the camera sensor plane. A large collimating lens is used for the reference arm to generate quasi-plane beam. After collimating, the beam waist diameter of the reference should be larger than the size of the camera sensor.

\subsubsection{Off-axis digital holography}

\begin{figure}[!h]
    \centering
    \includegraphics[width=6in]{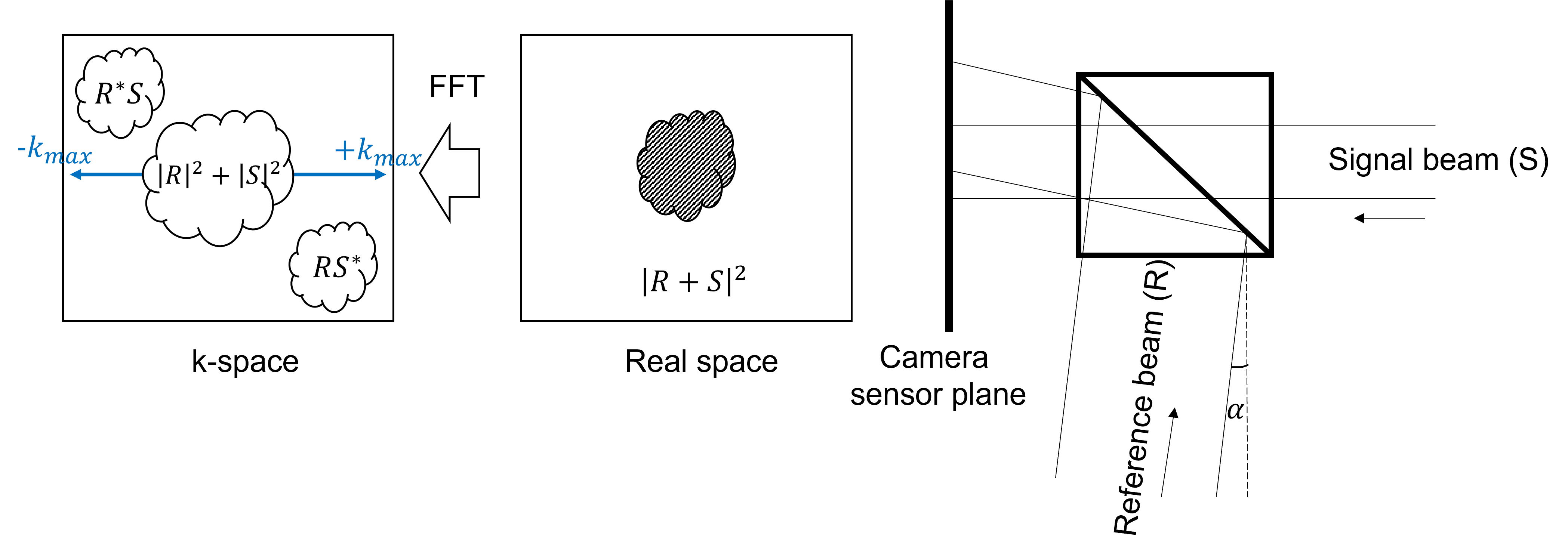}
    \caption{Schematic of off-axis digital holography.}
    \label{mplc_holography}
\end{figure}

Off-axis digital holography was used to measure the complex field and further the complete mode transfer matrix of the mode multiplexer/demultiplexer. Digital holography allows us to do the alignment of optical fields and modes decomposition in the computer, without introducing possible errors of the characterization setup itself \cite{fontaine2019laguerre,shaked2020off,mazur2019characterization}. 

Fundamentals of off-axis digital holography can be found in Dr. Joel Carpenter's digHolo library shared on GitHub and YouTube \cite{carpenter2022digholo}. Chapter 3 of Dr. Heide's dissertation is another good reference for the basics of off-axis digital holography \cite{van2022space}. Ref\cite{verrier2011off} talks about some practical considerations when doing off-axis digital holography.

The schematic of off-axis digital holography is shown in Fig.\ref{mplc_holography}. The complex field of the signal beam can be extracted from the 1st order in the Fourier domain. Ideally the lengths of the signal arm and reference arm should be matched and a highly coherent laser is used to increase the contrast of the fringes. The signal and reference fields should be mutually coherent during the exposure of the camera so that their relative phase is stable when capturing a frame. The maximum tilt angle $\alpha$ of the reference beam is limited by the pixel size of the camera $\Delta x$ and the wavelength. As we discussed the in chapter 2.1.2, the spatial frequency range is $-1/(2\Delta x) $ to $1/(2\Delta x)$, so that $k_{max}=2\pi max(v_x)=\pi/\Delta x$. For paraxial beam, we have 
\begin{equation}
\alpha _{max} \approx sin\alpha _{max} = k_{max}/k_0 = \frac{\lambda}{2\Delta x}
\end{equation}

\begin{figure}[!t]
    \centering
    \includegraphics[width=5in]{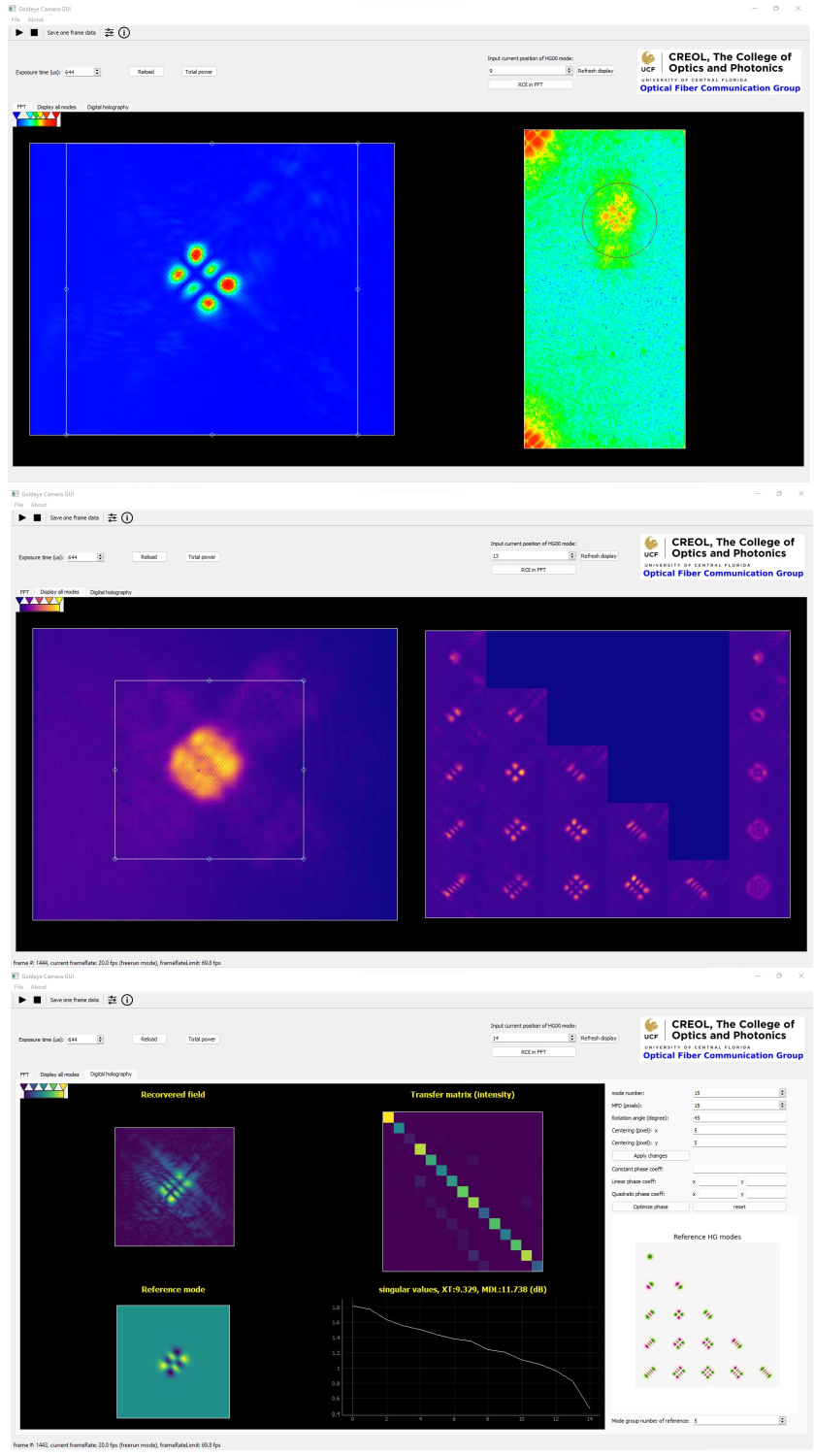}
    \caption{A home-made Python GUI for MPLC characterization.}
    \label{mplc_gui}
\end{figure}

For 1550 nm and a pixel size of 30$\times$30 $\mu m$ of the Goldeye camera (Allied Vision, Goldeye G-008 SWIR TEC1), the maximum angle is about 1.48 degree. If the tilt angle is larger than this value, there will be more than one interference fringes inside one camera pixel which causes aliasing, as a result the fringe contrast reduces and the recovered field is not accurate.

In real experiment, misalignment of the Fourier lens, tilt of the camera sensor, or the
quasi-plane feature of the big reference beam introduce phase errors to the system. The linear (tip and tilt) and quadratic (defocus) phase errors, and the centering in $x$ and $y$ of the recovered field, as well as the beam size and rotation angle of the reference modes should be optimized before calculating the mode transfer matrix. Linear phase errors can be compensated by centering the beam in the Fourier domain ($k$-space), while centering in $x$ and $y$ can be done by removing the linear phase in $k$-space. The defocus can be compensated by multiplying a thin lens function in real space. 

To assist the MPLC alignment and characterization, a graphical user interface (GUI) was developed in the lab  as shown in Fig.\ref{mplc_gui}. The top figure shows the raw image on the left and the real FFT on the right, with a circle ROI (region of interest) filter. The middle figure shows the summation of all raw images on the left, and all output modes corresponding to each input port on the right. This is enabled by synchronizing the optical switch and the camera using an external trigger. Modes within the same row belong to the same mode group and are degenerate. The last column sums the intensity of all modes of the same group and shows shapes of concentric circles as mentioned in \cite{fontaine2018packaged}. The bottom figure shows the intensity of the recovered field, the ideal mode to do overlap integral, the mode coupling matrix, and the singular values, XT and MDL values. If the data is processed fast enough, the mode coupling matrix, XT, and MDL values can be updated almost in real time, which is very helpful to align the MPLC. PyQt5 and PyQtGraph \cite{PyQtGraphwebsite} packages were used for the GUI geometric layout. Multi-threading programming could be applied to improve the processing speed of the GUI.

\subsubsection{Make a multimode collimator}

\begin{figure}
    \centering
    \includegraphics[width=5.6in]{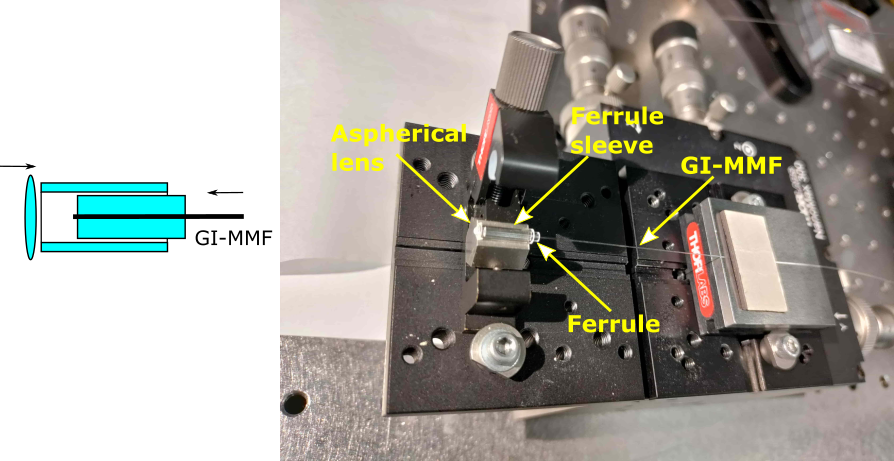}
    \caption{Make a multimode collimator for the MPLC.}
    \label{mplc_MMFcollimaotr}
\end{figure}

A multimode collimator is necessary if we want to couple the free space output of MPLC into a MMF \cite{jung2020compact,fontaine2022photonic}. Since HG modes are the eigen modes of a medium with parabolic index profile, the GI-MMF is usually used. As shown in Fig.\ref{mplc_MMFcollimaotr}, first we insert the cleaved MMF into a ferrule and glue it. Then we glue an aspherical lens or a GRIN lens to the ferrule sleeve. Finally, the glued ferrule is carefully inserted into the sleeve and by adjust the spacing to the lens to make sure it's focused to the far field, i.e., collimated. The designed output beam size of the MPLC determines which lens to use. It's important to make sure the collimated beam waist is very close to the lens surface so that we can glue the multimode collimator on the same glass base holding the MPLC by matching their beam waists. 

\subsubsection{Multi-step binary lithography}

\begin{figure}
    \centering
    \includegraphics[width=5.6 in]{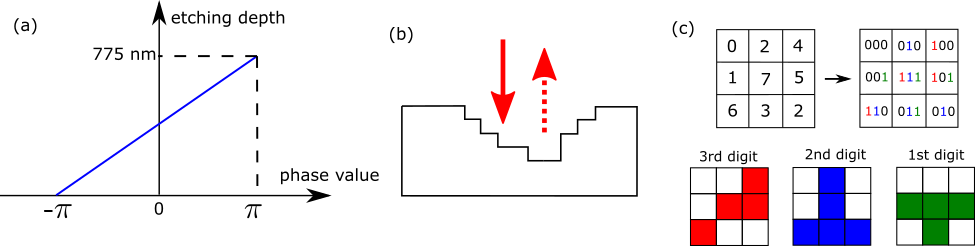}
    \caption{(a) Mapping relation between wrapped phase and etching depth for a reflective MPLC at a center wavelength of 1550 nm. (b) Cross-section of a wafer after etching using the multi-step binary lithography. (c) Principles of the multi-step binary lithography using the example of $2^3=8$.} 
    \label{mplc_lithography}
\end{figure}

Multi-step binary lithography (also known as gray-scale lithography) is the technique to fabricate phase masks on the fused silica wafer. Fig.\ref{mplc_lithography}(a) shows the mapping relation between etching depth and the wrapped phase of the mask when the center wavelength is 1550 nm.  Fig.\ref{mplc_lithography}(b) schematically shows the cross-section of the etched wafer with digitized steps. Reflected beam accumulates different phases at different pixels due to the phase retardation encountered at that pixel. Note that for a centering wavelength of 1550 nm, maximum etching depth is 1550/2 = 775 nm because it works in the reflective mode. In Fig.\ref{mplc_lithography}(c) we illustrate the principle of multi-step binary lithography. Taking $2^3 = 8$ for example, 3 etching steps give 8 discrete phase levels between 0 and 7,  which are randomly distributed in a 3$\times$3 box. Next we convert each number to the binary format and number "1" in a certain digit means it's going to be etched. Red, blue and green color is used to show the etching profile corresponding to a certain digit respectively.
3 layers of photomasks are required for this fabrication. Each layer has a different etching depth related to the "weight" of that binary format. 

Besides varying the physical thickness to generate gray-scale phase profiles, \cite{divliansky2019wavefront} also mentioned using a photosensitive material together with a digital micromirror device (DMD) to create arbitrary phase profiles. The spatial shape of the phase pattern is controlled by the DMD, and the phase value at each pixel is determined by the exposure time of micromirrors in the "on" position. 

\subsubsection{Align a 15-mode MPLC with the LCoS and fabricated phase masks}

\begin{figure}
    \centering
    \includegraphics[width=6 in]{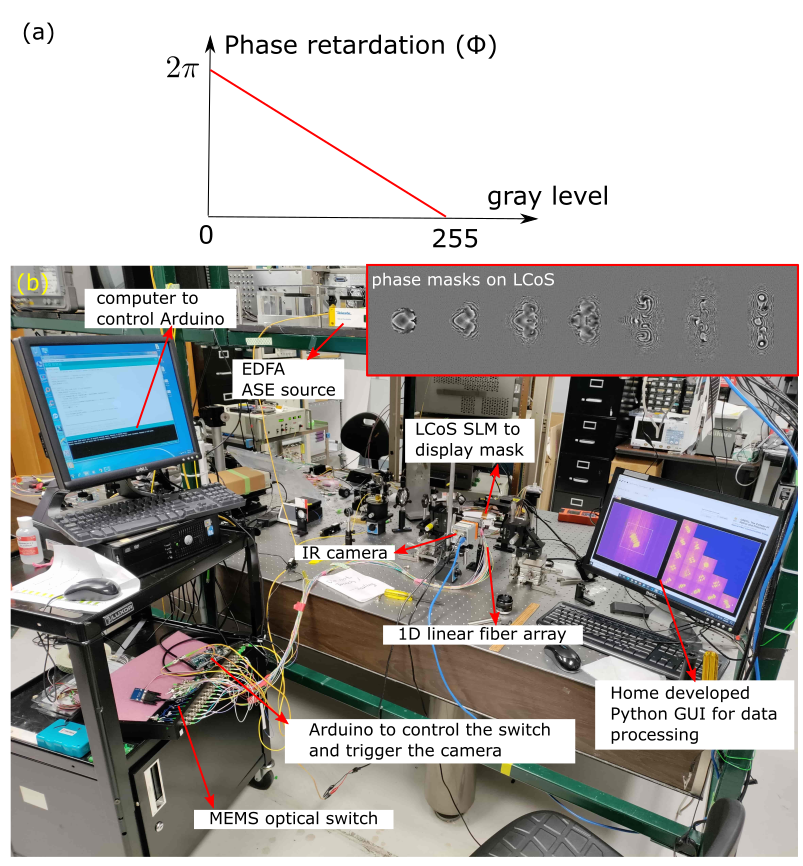}
    \caption{(a) The Holoeye PLUTO-2-TELCO-059 Phase Only SLM has a reversed phase response. (b)Align a 15-mode MPLC with phase masks on the phase-only LCoS. Inset shows the phase mask on the LCoS.}
    \label{mplc_15mode_lcos}
\end{figure}

\begin{figure}
    \centering
    \includegraphics[width=6 in]{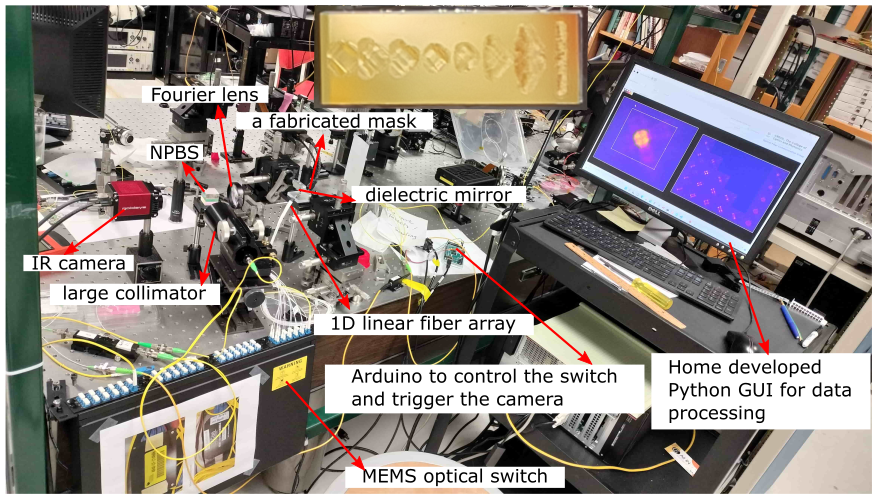}
    \caption{Align a 15-mode MPLC with fabricated phase masks, in a off-axis digital holography setup. Inset shows a photo of the fabricated phase mask, coated with gold. NPBS: Non-polarization beam splitter.}
    \label{mplc_15mode_fab}
\end{figure}

In this section we show the experimental setup for the alignment of a 15-mode MPLC using a LCoS and a fabricated phase mask, shown in Fig.\ref{mplc_15mode_lcos} and Fig.\ref{mplc_15mode_fab}, respectively. The Holoeye PLUTO-2-TELCO-059 Phase Only LCoS was used for the experiment. It should be noted that the phase response is wavelength dependent. For certain LCoS (like the one used in our experiment) max retardation is present in off state. After calibration the phase retaration is in a decreasing linear manner when addressing gray level from 0-255, as shown in Fig.\ref{mplc_15mode_lcos} (a). And you need to reverse the phase mask before putting it on the LCoS. We should also pay attention to the fringing field effect of LCoS \cite{persson2012reducing,moser2019model} that introduces inter-pixel crosstalk and gives a larger MDL. This phase-only LCoS is polarization sensitive and only gives pure phase modulation if the input beam has correct polarization, otherwise there will be phase-amplitude crosstalk \cite{zhang2014fundamentals}. \cite{ngcobo2013digital} briefly reviewed a method to implement amplitude modulation on a phase-only LCoS.

\section{Conclusion}
Theoretically, MPLC can be used to realize any unitary conversion using a set of phase masks. However, only certain transforms (right triangle array of spots to Hermite-Gaussian modes) was found that only requires a small number of phase masks. The trade-off between the number of phase masks and the performance (bandwidth, MDL, crosstalk, etc.) is a question that needs serious attention at the beginning of every design. In reality, it's ideal to reduce the number of phase masks both for easy alignment/fabrication and for reducing the total loss of the system. The lithography-defined phase masks can be fabricated at a large scale with a reasonable cost, but will also introduce scattering loss due to the pixelized structures. In theory the loss can be well below 1 dB, but current MPLC devices has a loss between 3 $\sim$ 5 dB due to scattering \cite{fontaine2022photonic}. In the future, we believe new fabrication techniques will bring the experimental loss down to theoretical level and MPLC-based devices will be widely deployed. 

\clearpage

\bibliographystyle{ieeetr}
\bibliography{refs.bib}

\end{document}